# A simulation engine to support production scheduling using genetics-based machine learning

*Tamaki, H., Kryssanov, V.V., and Kitamura, S.*
*Faculty of Engineering, Kobe University*
*Rokko-dai, Nada-ku, Kobe 657-8501, Japan*
*Tel./Fax: +81-078-803-6102,*
*e-mail:kryssanov@ziong.cs.kobe-u.ac.jp*

**Abstract**
The ever higher complexity of manufacturing systems, continually shortening life cycles of products and their increasing variety, as well as the unstable market situation of the recent years require introducing grater flexibility and responsiveness to manufacturing processes. From this perspective, one of the critical manufacturing tasks, which traditionally attract significant attention in both academia and the industry, but which have no satisfactory universal solution, is production scheduling. This paper proposes an approach based on genetics-based machine learning (GBML) to treat the problem of flow shop scheduling. By the approach, a set of scheduling rules is represented as an individual of genetic algorithms, and the fitness of the individual is estimated based on the makespan of the schedule generated by using the rule-set. A concept of the interactive software environment consisting of a simulator and a GBML simulation engine is introduced to support human decision-making during scheduling. A pilot study is underway to evaluate the performance of the GBML technique in comparison with other methods (such as Johnson's algorithm and simulated annealing) while completing test examples.

1   INTRODUCTION

Recently, there is a widening belief that by applying simulation systems, it becomes possible to cope with the increasing technical, structural and organizational complexity of modern manufacturing enterprises while efficiently arranging production and administrative processes throughout the product life cycle. Manufacturing simulation systems help to realize and optimize deeper the structure and properties of professional activities underlying technological processes, find a satisfying ('good') solution for a problem among the variety of feasible alternatives, and predict and analyze potential consequences (immediate as well as distant) of a candidate decision made concerning any of the product life cycle stages. One of the most promising manufacturing activities to employ simulation techniques and tools is production scheduling.

Scheduling is well recognized as a complex task, which requires taking into account multiple factors at any time in the shop floor, where scheduling occurs as a resource allocation problem subject to meet dynamically changing resource constraints. Due to its complexity and importance for actual manufacturing, scheduling traditionally attracts much research interest and in the recent years, a number of computational approaches, computer-based methods and systems have been proposed to facilitate and automate this activity (see Chrétienne *et al.*, 1995; Brucker, 1995). The developed approaches can roughly be classified into three categories: deterministic (e.g. the branch and bound method), search-based (genetic algorithms and the like), and hybrid (e.g. neural networks controlling genetic algorithms). Although methods of the first group can successfully be applied to obtain optimal schedules for small-sized problems, they are not applicable to larger problems as the computation cost increases exponentially with the growth of the problem complexity. Search-based approaches can be used to obtain nearly optimal solutions with a reasonable computational cost in most cases, but there is a problem in determining the right values for the parameters of the algorithms utilized. The latter usually requires a great deal of experimenting to find parameters appropriate for the given problem that makes it difficult to promptly react to the environmental changes. Hybrid methods seem to be free of the aforementioned bottlenecks, but they are usually difficult and expensive to implement. Therefore, other approaches need to be found and explored, which would allow for solving manifold scheduling problems.

In the presented study, a new simulation technique based on genetics-based machine learning is proposed and applied to support decision-making during completing schedules for a flow shop environment with finite and infinite buffers. A decision-making support tool is developed, and a pilot study is made to validate the approach.

## 2 PROBLEM DEFINITION

We will consider a production system environment consisting of $m$ machines $M_i$, $i=1,\ldots,m$, assigned to accomplish $n$ jobs $J_j$, $j=1,\ldots,n$. Each job $J_j$ includes $n_j$ operations $O_{jk}$, $k=1,\ldots,n_j$. It is assumed that exactly one machine is assigned to every operation with the processing time $p_{jk}$, there are no machine breakdowns, and the jobs are available at time zero and have sequence-dependent setup times on each machine. The system principal parameters, such as processing times and setup times, are supposed to be known with certainty for the given interval of time.

The flow shop scheduling problem in the environment can then be formulated as follows: to find a sequence of jobs that satisfies certain optimization criteria and environmental constraints. This problem belongs to combinatorial optimization problems and in many practically important cases, it is NP-hard (Blazewicz *et al.*, 1996).

## 3 FORMALIZATION OF THE SCHEDULING PROBLEM

A priority-based scheme of scheduling is utilized that includes the following steps:

1. Create a list of jobs to be processed.
2. If there is no job in the list, then go to Step 5.
3. Calculate priorities of the jobs to be processed.
4. Select the job with the highest priority and calculate the start time of each operation of the job. Eliminate the job from the list of jobs and go to Step 2.
5. Terminate with a complete schedule.

The case of non-delay schedules (i.e. when the operations of a job should start as early as possible) is considered. The priority $\alpha_j$ of a job $J_j$ is calculated as $\alpha_j = \sum w_i a_{ji}$, where $i=1,\ldots,n_A$, $a_{ji}$ is the $i$-th attribute (or status) of the job $J_j$, $w_i$ is a weight value, and $n_A$ is the number of attributes. Then, a scheduling rule can be represented in the form of a weight vector $(w_1,\ldots,w_{n_A})$.

It is assumed that the state space of the production environment can be decomposed into $n_S$ subsets $S_1,\ldots,S_{n_S}$ in such a way that each subset corresponds to a distinct (nearly optimal) scheduling rule-set: if $s \in S_k$ then assign $(w_{k1},\ldots,w_{kn_A})$, $k=1,\ldots,n_S$, where $s$ is a vector characterizing the system current state. Below, a simulation technique is described that allows for appropriately adjusting the weights and obtaining nearly optimal schedules.

## 4 GBML APPROACH

We apply a genetics-based machine learning (GBML) technique (see Goldberg, 1989) to calculate weight vectors (scheduling rules). A rule-set is represented as an individual of a genetic algorithm. If a scheduling problem $H_i$, $i=1,\ldots,n_H$, where $n_H$ is the number of scheduling problems to be done, has been formulated, the algorithm consists of the following steps:

1. Set counter $t=1$. Randomly generate $N_p$ rule-sets and by numerically encoding these rule-sets, form an initial population $P(t)$.
2. If $t>N_g$, where $N_g$ is the number of generations to be produced, then go to Step 5.
3. Create schedules by applying each individual (rule-set) to the problem. Evaluate the schedules and calculate the fitness value for every individual.
4. Generate a new population $P(t+1)$ of the next generation by applying genetic (recombination, mutation, and reproduction) operations to the population $P(t)$. Set $t=t+1$, go to Step 2.
5. Select the best-so-far individual and terminate.

An individual is represented through encoding the corresponding rule-set as a linear array of integers so that an attribute weight is ranged as an integer. The fitness $F_i$ of an individual $i$ is calculated as a weighed mean of objective function values estimated for different instances of the scheduling problem:

$F_i = max[0, m - \Sigma(o_{ij} / \bar{o}_j)]$, where $j=1,\ldots,n_H$, m is a constant, $o_{ij}$ is the objective function value obtained by applying the individual $i$ to the problem $H_j$, and $\bar{o}_j$ is the calculated mean value of the objective function values for the problem $H_j$.

The genetic algorithm drives the evolution of the population, performing mutation, recombination, selection, and reproduction. For the selection and reproduction operations, the remainder stochastic sampling with replacement method (Michalewicz, 1992) has been adopted, and the elitist strategy (Goldberg, 1989) has been used. More details about the genetic algorithm employed in the research can be found in (Tamaki *et al.*, 1996).

## 5  A PILOT STUDY

To explore the applicability of the proposed approach to the production scheduling problem, a decision-making support tool has been developed in our study (see Figure 1). The underlying idea of the system is that an automatic scheduler cannot address all the aspects related to the dynamic settings of a production system due to the system complexity, the unpredictability of the environment, and general difficulties in obtaining relevant information and data. The software tool is therefore needed to support human decision-making rather than to replace the operator.

There are two main components of the developed system: a GBML engine and a simulator. The former is to drive generation of scheduling rule-sets, and the latter is to evaluate these rule-sets based on the environmental data and optimization criteria specified by the operator, and to calculate schedules. The operator can run the simulator and, having assigned a state vector, determine a (set of) schedule(s), which is (nearly) optimal for the given state of the production system. The decision-making support tool can also be used in an interactive mode to perform exploration and optimization of completed schedules by using the genetic algorithm.

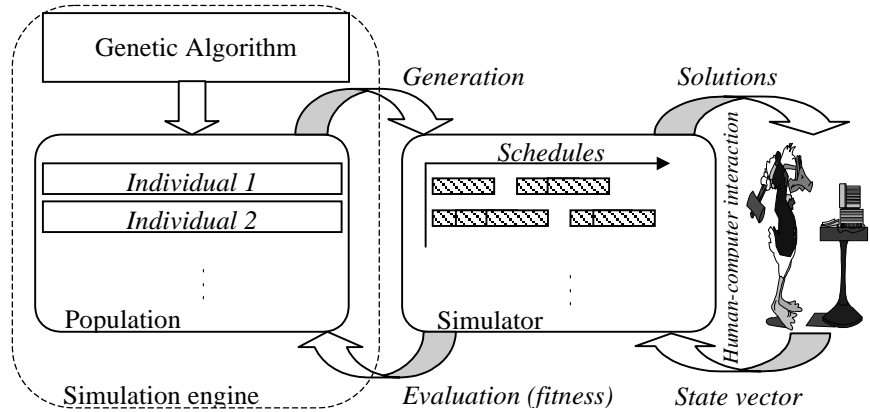

**Figure 1.** Software environment.

A flow shop scheduling problem with 2 machines, an intermediate (work-in-process) buffer, and 50 jobs ($m=2$, $n=50$, and $\forall j$, $n_j=50$) has been considered, and the criterion of the minimal makespan ($max\ c_j$) $\rightarrow min$, where $c_j$ is the completion time of $J_j$, has been applied to calculate schedules. 40 examples of scheduling problems have been prepared by randomly determining the processing time $p_{jk}$ and varying the capacity of the buffer $k$: $H_{ik}$, $i=1,\ldots,10$ and $k \in \{1, 3, 5, \infty\}$. The state space of the system has been divided into 8 specific subsets ($n_S=8$) defined for the given environment by a subject matter expert as it is shown in Figure 2. Two attributes have been specified for each job ($n_A=2$): $a_{j1}$ – the processing time of the job $J_j$ on Machine 1, and $a_{j2}$ – the processing time of the job $J_j$ on Machine 2.

Ten trials (simulations) have been made with each example. Table 1 gives results of the simulations in comparison with results of the application of other approaches to the same scheduling problems. (Mean values of the objective functions calculated through the trials have been estimated for the comparison).

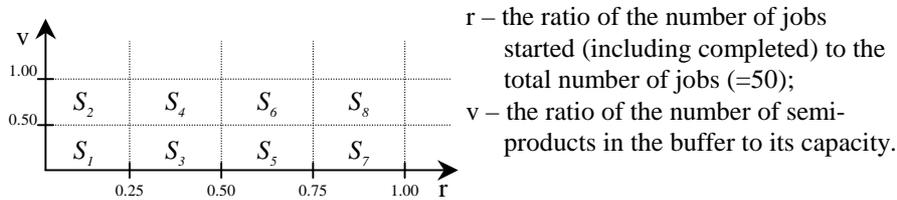

r – the ratio of the number of jobs started (including completed) to the total number of jobs (=50);
v – the ratio of the number of semi-products in the buffer to its capacity.

**Figure 2.** The state space decomposition.

Table 1 Results of trials of scheduling by GBML, Johnson's Algorithm, and Simulated Annealing and their comparison

| No. of the example, $i$ | Buffer capacity, $k$ | The mean value of the objective functions | | | I/II | I/III | II/III |
|---|---|---|---|---|---|---|---|
| | | GBML | Johnson's algorithm | Simulated annealing | | | |
| | | I | II | III | | | |
| 1  | ∞ | 243   | 243 | -   | 1.000 |       |       |
| 2  |   | 299   | 299 | -   | 1.000 |       |       |
| 3  |   | 286   | 286 | -   | 1.000 |       |       |
| 4  |   | 290   | 290 | -   | 1.000 |       |       |
| 5  |   | 349   | 349 | -   | 1.000 |       |       |
| 6  |   | 325   | 325 | -   | 1.000 |       |       |
| 7  |   | 252   | 252 | -   | 1.000 |       |       |
| 8  |   | 298   | 298 | -   | 1.000 |       |       |
| 9  |   | 305   | 305 | -   | 1.000 |       |       |
| 10 |   | 256   | 256 | -   | 1.000 |       |       |
| 1  | 1 | 243   | 261 | 243 |       | 1.000 | 1.074 |
| 2  |   | 299   | 313 | 299 |       | 1.000 | 1.047 |
| 3  |   | 286   | 398 | 286 |       | 1.000 | 1.392 |
| 4  |   | 294.2 | 372 | 290 |       | 1.014 | 1.283 |
| 5  |   | 349   | 370 | 349 |       | 1.000 | 1.060 |
| 6  |   | 325.4 | 356 | 325 |       | 1.001 | 1.095 |
| 7  |   | 252   | 274 | 252 |       | 1.000 | 1.087 |
| 8  |   | 298   | 313 | 298 |       | 1.000 | 1.050 |
| 9  |   | 305   | 334 | 305 |       | 1.000 | 1.095 |
| 10 |   | 260   | 305 | 256 |       | 1.016 | 1.191 |
| 1  | 3 | 243   | 247 | 243 |       | 1.000 | 1.016 |
| 2  |   | 299   | 300 | 299 |       | 1.000 | 1.003 |
| 3  |   | 286   | 380 | 286 |       | 1.000 | 1.329 |
| 4  |   | 290.2 | 356 | 290 |       | 1.014 | 1.228 |
| 5  |   | 349   | 354 | 349 |       | 1.000 | 1.014 |
| 6  |   | 325   | 343 | 325 |       | 1.000 | 1.055 |
| 7  |   | 252   | 265 | 252 |       | 1.000 | 1.052 |
| 8  |   | 298   | 304 | 298 |       | 1.000 | 1.026 |
| 9  |   | 305   | 318 | 305 |       | 1.000 | 1.043 |
| 10 |   | 256   | 291 | 256 |       | 1.000 | 1.137 |
| 1  | 5 | 243   | 243 | 243 |       | 1.000 | 1.000 |
| 2  |   | 299   | 299 | 299 |       | 1.000 | 1.000 |
| 3  |   | 286   | 363 | 286 |       | 1.000 | 1.269 |
| 4  |   | 290.2 | 342 | 290 |       | 1.014 | 1.179 |
| 5  |   | 349   | 349 | 349 |       | 1.000 | 1.000 |
| 6  |   | 325   | 329 | 325 |       | 1.000 | 1.012 |
| 7  |   | 252   | 253 | 252 |       | 1.000 | 1.004 |
| 8  |   | 298   | 304 | 298 |       | 1.000 | 1.020 |
| 9  |   | 305   | 308 | 305 |       | 1.000 | 1.010 |
| 10 |   | 256   | 278 | 256 |       | 1.000 | 1.086 |

Johnson's algorithm (Johnson, 1954), which guarantees generation of the optimal schedule for the environments with $m=2$ and infinite buffer capacity, and the simulated annealing technique, which demonstrated the best performance in solving scheduling problems similar to the tests (Tamaki *et al.*, 1993), have been used to evaluate the effectiveness of the rules evolved out of the initial population by using GBML.

## 6   DISCUSSION AND CONCLUSIONS

The experimental results represented in Table 1 clearly indicate that the GBML approach ensures generating optimal schedules in the case of the flow shop environment with infinite buffer capacity, and nearly optimal – in the case of the environment with the fixed buffer capacities. Overall, the efficiency and effectiveness of applying GBML to the considered scheduling problems is high. Results of other experiments reported in (Tamaki *at al.*, 1998) showed that schedules obtained by applying GBML are potentially robust, and changing values of the state vector within the ranges specified by the system state decomposition does not significantly affect the performance of the rule-sets.

Furthermore, our experience has been that the implementation of the software environment did not take long time and required reasonable resources, and that the proposed software offers a convenient and natural (from the standpoint of the operator's logic of decision-making) information support structure for the interactive development of schedules in dynamic flow shop settings.

The main difficulties encountered in using the GBML approach are determining a state vector that properly describes the production system and assigning the state space decomposition for the industrial-size scheduling problems. These, as well as finding strategies for effective human-computer interaction remain for future research.

Thus, in the presented paper, a new simulation approach to support the process of the creation of schedules has been proposed. The approach is generative but not adaptive, and it employs a genetics-based machine learning technique to build feasible schedules. To explore the approach applicability, a decision-making support tool has been developed and a pilot study has been made, calculating schedules for a flow shop production environment. The study results confirmed the efficiency of applying the GBML technique to scheduling problems.

## 7   ACKNOWLEDGEMENTS

This work relates to research supported by the Japan Society for the Promotion of Science (JSPS); the 'Methodology of Emergent Synthesis' Project (No. 96P00702) under the Program 'Research for the Future.'

## 8 REFERENCES

Blazewicz, J., Ecker, K., Pesch, E., Schmidt, D., and Weglarz, J. (1996). Scheduling Computer and Manufacturing Processes. Berlin: Springer-Verlag

Brucker, P. (1995). Scheduling Algorithms. Berlin: Springer-Verlag

Chrétienne, P., Coffman, E. G., Lenstra, J. K., and Liu, Z. (1995). Scheduling Theory and Its Applications. Chichester: John Wiley & Sons

Goldberg, D.E. (1989). Genetic Algorithms in Search, Optimization, and Machine Learning. Reading, MA: Addison-Wesley

Johnson, S. M. (1954). Optimal two- and three-stage production schedules with setup times included. *Naval Research Logistics Quarterly*, 1, 61-68

Tamaki, H., Hasegava, Y., Kozasa, J., and Araki, M. (1993). Scheduling in Plastics Forming Plant: A Binary Representation Approach, In: *Proceedings of the $32^{nd}$ CDC*, pp.3845-3850

Tamaki, H., Ochi, M., and Araki, M. (1996) Application of Genetics-Based Machine Learning to Production Scheduling, In: *Proceedings of the 1996 Japan-USA Symposium on Flexible Automation*, pp. 1221-1224

## 9 BIOGRAPHY

Hisashi Tamaki, Dr. Eng., is an associate professor at the Department of Electrical and Electronics Engineering, the Faculty of Engineering of Kobe University, Japan.

Victor V. Kryssanov, Ph. D., is a research associate at the Department of Computer and Systems Engineering, the Faculty of Engineering of Kobe University, Japan.

Shinzo Kitamura, Ph. D., is a professor at the Department of Computer and Systems Engineering and the dean of the Faculty of Engineering of Kobe University, Japan.